# Image Magnification Network for Vessel Segmentation in OCTA Images

Mingchao Li[1], Yerui Chen[1], Weiwei Zhang[2] and Qiang Chen[1]

[1] School of Computer Science and Engineering, Nanjing University of Science and Technology, Nanjing 210094, China
[2] Department of Ophthalmology, The First Affiliated Hospital with Nanjing Medical University, Nanjing 210094, China
`Chen2qiang@njust.edu.cn`

**Abstract.** Optical coherence tomography angiography (OCTA) is a novel non-invasive imaging modality that allows micron-level resolution to visualize the retinal microvasculature. The retinal vessel segmentation in OCTA images is still an open problem, and especially the thin and dense structure of the capillary plexus is an important challenge of this problem. In this work, we propose a novel image magnification network (IMN) for vessel segmentation in OCTA images. Contrary to the U-Net structure with a down-sampling encoder and up-sampling decoder, the proposed IMN adopts the design of up-sampling encoding and then down-sampling decoding. This design is to capture more image details and reduce the omission of thin-and-small structures. The experimental results on three open OCTA datasets show that the proposed IMN with an average dice score of 90.2% achieves the best performance in vessel segmentation of OCTA images. Besides, we also demonstrate the superior performance of IMN in cross-field image vessel segmentation and vessel skeleton extraction.

**Keywords:** OCTA, image magnification network, retina, vessel segmentation.

## 1 Introduction

Optical coherence tomography angiography (OCTA) is a relatively novel imaging modality for the micron-level imaging of the retinal microvasculature [1]. Compared to alternative angiographies, such as fluorescein angiography, it is fast, non-invasive and allows to provide angiographic images at different retinal depths. With these advantages, OCTA has been used to evaluate several ophthalmologic diseases, such as age-related macular degeneration (AMD) [2], diabetic retinopathy (DR) [3], artery and vein occlusions [4], and glaucoma [5], etc. A more recent study demonstrated that the morphological changes of microvascular calculated from OCTA images are related to Alzheimer's Disease and Mild Cognitive Impairment [6]. Thus, quantitative analysis of OCTA images is of great value for the diagnosis of the related diseases.

Quantitative phenotypes in OCTA image such as vessel density (VD) [7], vessel tortuosity (VT) [8], fractal dimension (FD) [9] rely on the segmented vessel masks. It is extremely time-consuming and laborious to draw these masks manually, so it is necessary to design automatic segmentation approaches for OCTA images. But there exist the following challenges (as shown in Figure 1): (a) OCTA images contain a lot of high-value noise. (b) Some vascular signals are weak due to turbid refractive medium. (c) The capillary plexus is densely structured which is difficult to be identified. (d) The capillary plexus is very thin, most of which are only 1 pixel wide.

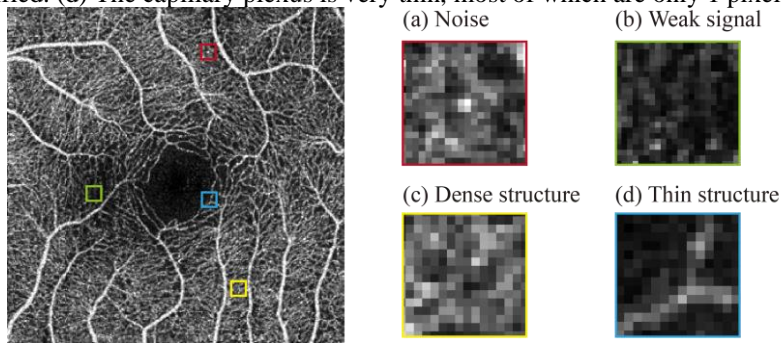

**Fig. 1.** An example of OCTA image projected within inner retina and its segmentation challenges.

Several approaches have been used for vessel segmentation in OCTA images, including threshold-based methods [10, 11], filter-based methods [12-16], and deep learning-based methods [17-22]. Threshold-based methods use global threshold or adaptive threshold to directly binarize OCTA images. The binary masks obtained by the threshold-based methods retain a lot of noise and many weak signal vessels are lost. Filter-based methods can suppress the noise in OCTA images to better visualize the microvasculature, but their quantitative performances are even worse than that of the threshold-based method as shown in Table 1. Besides, their performances are sensitive to the filter parameters. Deep learning-based methods have been used for vessel segmentation in OCTA images and obtained more excellent performance than filter-based methods. Mou et al. [18] introduced CS-Net, adding a spatial attention module and a channel attention module to the U-Net structure [19] to extract curvilinear structures from three biomedical imaging modalities including OCTA. Pissas et al. [20] recursively connected multiple U-Net to obtain refined predictions. More recently, Ma et al. [21] used two U-Net structures to segment pixel-level and centerline-level vessels and fuse them to obtain the final segmentation result.

The above deep learning-based methods use U-Net as the base structure to segment the retinal vessels. However, we find that the U-Net structure has limited detail protection ability, and it is easy to miss weak signal vessels and thin-and-small capillary. In this work, we propose an image magnification network (IMN), which adopts the design of up-sampling encoding and then down-sampling decoding, like a magnifying glass. This design is to capture more image details and reduce the omission of thin-and-small structures. This paper makes four contributions: (1) IMN is a novel end-to-end structure for vessel segmentation in OCTA images. (2) IMN has better detail protection performance than U-Net. (3) Experimental results on three open datasets demonstrate the state-of-the-art performance. (4) We explored the performance of the IMN model on cross-FOV images and its application in vessel skeleton segmentation.

## 2      Network Architecture

The structure of our proposed IMN is illustrated in Figure 2(b). Compared with the typical U-Net structure [19] (Figure 2(a)), there is a simple change, that is, the proposed IMN consists of an up-sampled encoding path and a down-sampled decoding path, while the U-Net is composed of a down-sampled encoding path and an up-sampled decoding path. This structural difference is determined by the particularity of the vessel segmentation task in OCTA images. In our practice, this simple change is very useful for identifying thin-and-small vessels in OCTA images.

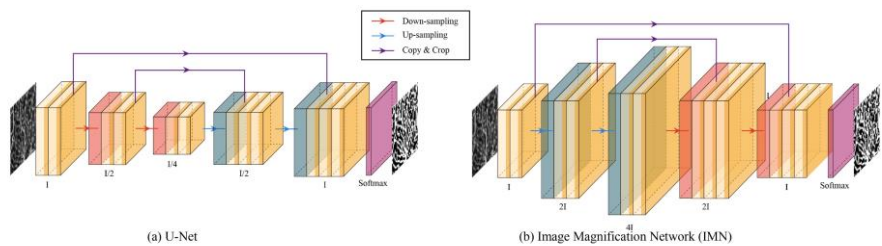

**Fig. 2.** Architectures of the typical U-Net (a) and Image Magnification Network (b).

**Our key insight is using the up-sampling encoder to maintain image details.** The down-sampling encoder in U-Net compresses the image to extract the high-level features, thereby identifying the specified structure in the complex content images. However, the content in the OCTA image is relatively simple, including only blood flow signal and noise signal. The high-level features extracted by the encoder in U-Net have a limited gain in vessel segmentation tasks. At the same time, the detailed information of the image will be weakened during the down-sampling operation and the convolution operation, resulting in incomplete segmentation of thin-and-small capillary structures (as shown in Figure 3, yellow arrow). On the contrary, our proposed IMN contains an up-sampled encoding path and a down-sampled decoding path, which is to magnify the image details like using a magnifying glass, preventing the detailed information from being weakened and lost, thereby obtaining a more complete capillary structure.

In IMN, the encoding path consists of the repeated structure of two 3×3 convolution layers and one up-sampling layer. Each convolution operator follows a batch normal (BN) layer and a rectified linear unit (ReLU). Transposed convolution with the kernel of 2×2 and the stride of 2×2 is used for up-sampling. The decoding path contains the repeated structure of two 3×3 convolution layers same as the encoding path and follows by a 2×2 max-pooling layer for down-sampling. We use copy-and-crop operation as the connection method to connect the selected layers of the encoding path and the decoding path. The channel number of the convolutional layers is set to 256, and the 256 feature vectors are mapped to 2 (number of classes) feature maps through an additional convolutional layer. We use cross entropy as loss function to supervise network training. The implementation based on PyTorch framework is available at **. (Publicly visible when the paper is accepted.)

## 3      Datasets and Evaluation Metrics

We evaluated the proposed method on the three recently published OCTA datasets: an OCTA segmentation study[1] [23] (called OCTA-SS in this paper), OCTA-500[2] [24], and ROSE[3] [21]. OCTA-SS provides the most detailed vessel annotation, which is used for quantitative comparison and evaluation. Regrettably, it has not provided the complete field of view (FOV) images, only 55 slices of the region of interest obtained from 3mm×3mm FOV images of 11 participants with and without family history of dementia. OCTA-500 provides three-dimensional OCTA data of 500 eyes (Disease types include AMD, DR, choroidal neovascularization (CNV), etc.) with two FOV (6mm×6mm,3mm×3mm) and complete projection images of different depths. We used OCTA-500 to evaluate the performance of the models trained by the OCTA-SS on the two complete FOV images. ROSE contains two sub-datasets, ROSE-1 and ROSE-2. In this paper our experiments are performed on ROSE-1, which consists of a total of 117 OCTA images from 39 subjects with and without Alzheimer's disease. The ROSE dataset provides centerline-level vessel annotations, which is also used to explore the performance of the proposed method in vessel skeleton segmentation.

Four commonly used evaluation metrics for image segmentation are calculated: Dice coefficient (Dice), accuracy (Acc), recall (Rec), and precision (Pre). They are denoted as:

$$Dice = 2TP/(2TP + FP + FN) \quad (1)$$
$$Acc = (TP + TN)/(TP + TN + FP + FN) \quad (2)$$
$$Rec = TP/(TP + FN) \quad (3)$$
$$Pre = TP/(TP + FP) \quad (4)$$

where TP is true positive, FP is false positive, FN is false negative, TN is true negative.

To evaluate the global quality of vessel segmentation, we also adopted the CAL metrics proposed in [25], which measures the connectivity, area and length of segmentation result. Furthermore, to evaluate the integrity of the vascular skeleton, we designed a skeleton recall rate (S-Rec), which uses the method [26] to extract the vascular skeleton and follow Formula 3 to calculate the recall rate of the skeleton image as S-Rec.

## 4      Experiments

The proposed IMN and the deep learning baselines we compared were implemented with PyTorch on 1 NVIDIA GeForce GTX 1080Ti GPU. We choose the Adam optimization with an initial learning rate of 0.0001, batch size of 1 and without weight decay. The threshold-based method and the filter-based based methods are run on the MATLAB platform, and all parameters are adjusted to the best performance through repeated experiments. In the OCTA-SS dataset, all images are cropped to 76×76. 30 images are used to train the model, and the remaining 25 images are used for testing and evaluation. ROSE dataset has a divided training set and a divided test set. The projection images in OCTA-500 are all regarded as the test set in this study.

---

[1] https://datashare.ed.ac.uk/handle/10283/3528

[2] https://ieee-dataport.org/open-access/octa-500

[3] https://imed.nimte.ac.cn/dataofrose.html

### 4.1 Comparison with Other Methods

Table 1. Performance of vessel segmentation methods on OCTA-SS dataset.

| Method | Dice | Acc | Rec | Pre | CAL | S-Rec |
|---|---|---|---|---|---|---|
| AT [27] | 0.8423 | 0.8879 | 0.9144 | 0.7922 | 0.7775 | 0.6699 |
| Frangi [12]+AT | 0.7967 | 0.8536 | 0.9237 | 0.7163 | 0.7100 | 0.6478 |
| SCIRD-TS [16]+AT | 0.7071 | 0.7704 | 0.8058 | 0.6446 | 0.6745 | 0.3912 |
| Gabor [13]+AT | 0.7688 | 0.8213 | 0.8643 | 0.7034 | 0.7609 | 0.5078 |
| CNN | 0.8943 | 0.9218 | 0.9197 | 0.8716 | 0.8991 | 0.7454 |
| U-Net [19] | 0.8958 | 0.9239 | 0.9105 | **0.8830** | 0.9038 | 0.7550 |
| CS-Net [18] | 0.8926 | 0.9213 | 0.9165 | 0.8711 | 0.8880 | 0.7563 |
| DUNet [28] | 0.8932 | 0.9224 | 0.9111 | 0.8776 | 0.8963 | 0.7528 |
| IMN | **0.9019** | **0.9268** | **0.9270** | 0.8794 | **0.9075** | **0.7740** |

We compared the proposed method with several baseline methods in the OCTA-SS dataset, including threshold-based methods, filter-based methods, and deep learning-based methods. For the threshold-based method, we adopted an adaptive threshold method (AT) [27]. For filter-based methods, we selected three well-known filters for vessel segmentation: Frangi [12], SCIRD-TS [16], and Gabor [13]. The binary segmentation results are obtained from the filtered image by using the adaptive threshold [27]. The implementation of the above methods can be found in [23]. For deep learning-based methods, U-Net [19] is used as an important baseline, and we also designed a CNN with 7 convolutional layers without up-sampling and down-sampling processes. The comparison of CNN, U-Net and IMN can be regarded as an ablation study to explore the effect of sampling. Furthermore, two state-of-the-art methods for blood vessel segmentation, CS-Net [18] and DUNet [28] are also used for comparison.

The segmentation performances of all methods according to the evaluation metrics are shown in Table 1. Several important properties can be summarized as: (1) The quantitative results of filter-based methods are mostly lower than those of the threshold-based method, and much lower than that of deep learning-based methods. (2) Our proposed IMN reaches the best results in Dice, Acc, Rec, CAL, and S-Rec. (3) The S-Rec score of IMN is significantly higher than other deep learning methods, indicating that the vessel skeleton of the IMN segmentation results is the most complete.

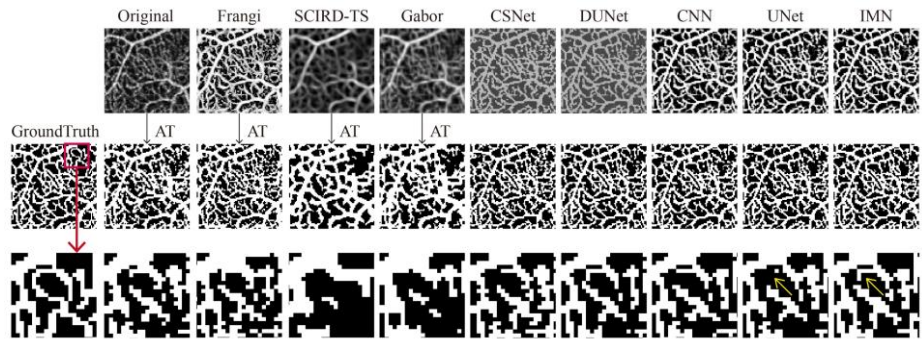

**Fig. 3.** Examples of vessel enhancement results (first row) and segmentation results (second row) between different methods. The third row shows the detailed segmentation results in the red box.

Figure 3 shows the segmentation results of an example with different methods. Interestingly, although the quantitative results of the filter-based methods are worse than those of the threshold-based method, they have a better visualization of capillaries and have the effect of vessel enhancement. The deep learning-based methods have achieved the segmentation performance closest to the ground truth. In particular, the proposed IMN has a better segmentation performance on thin-and-small blood vessels than U-Net (in Figure 3, yellow arrow), which hints at the reason for the highest S-Rec score of IMN.

By comparing CNN, U-Net and the proposed IMN in Table 1 and Figure 3, we find that the performance of U-Net and CNN are close in this task, and the performance of IMN is better. It indicates that the down-sampled encoding path in the U-Net structure has limited use for vessel segmentation in OCTA images, while the up-sampled encoding path in IMN stores more detailed information to prevent the loss of thin-and-small vessel structures.

### 4.2   Performance on Cross-FOV Images

OCTA-SS only provides slice images cropped from 3mm×3mm FOV images, which can only be used to observe the partial segmentation view. To explore the segmentation performance on the complete images and cross-FOV images, we run the trained model on the complete projection images with two FOV (6mm×6mm, 3mm×3mm) in the OCTA-500 dataset. Since the input size of the model is fixed, cropping and stitching are necessary. The final segmentation results are shown in Figure 4.

We find that there is no obvious error in the segmentation of the complete image, which shows that the method of training partial slices is feasible in this task, and there is no need to label and train the entire image. Interestingly, the segmentation model trained on the 3mm×3mm FOV images can also be used to segment the vessels of the 6mm×6mm FOV images, which shows that the deep learning-based methods have good portability in cross-FOV.

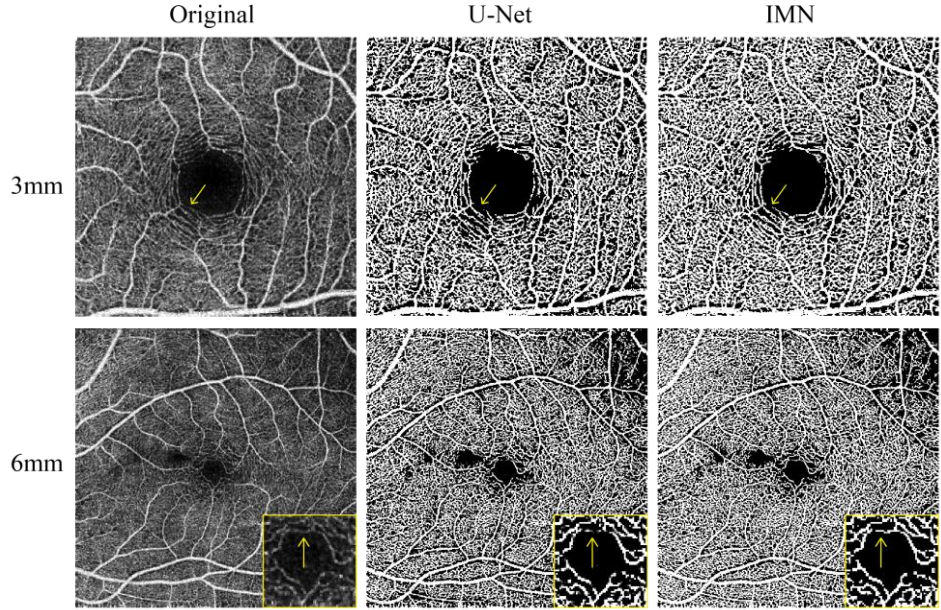

**Fig. 4.** The segmentation results on different FOV images by using U-Net and IMN. The first row is an example with 3mm×3mm FOV. The second row is an example with 6mm×6mm FOV.

We compared the results of the proposed IMN and U-Net. The segmentation performance of IMN still maintains more vessel details, and the thin-and-small vessels are better segmented, as shown by the yellow arrows in Figure 4. On the whole, the vessel segmentation results of IMN are more detailed and denser than U-Net, which once again proves the state-of-the-art performance of our proposed IMN.

### 4.3 Performance on Vessel Skeleton Segmentation

We have verified the excellent performance of IMN on the pixel-level vessel segmentation of OCTA images. Then, we explored vessel skeleton segmentation, which is one of the important research directions of vessel segmentation. The ROSE dataset contains the vessel skeleton labels as shown in figure 5(d). We trained U-Net and IMN models using these labels on the ROSE dataset, and the segmentation results are shown in Figures 5(b)(c). The vessel skeleton segmentation of IMN is slightly more complete than that of U-Net (Figure 5, red arrows). However, most of the capillary network skeleton has not been segmented, and the results are still far from the ground truth. Therefore, it is not suitable to segment the vessel skeleton directly using U-Net or IMN.

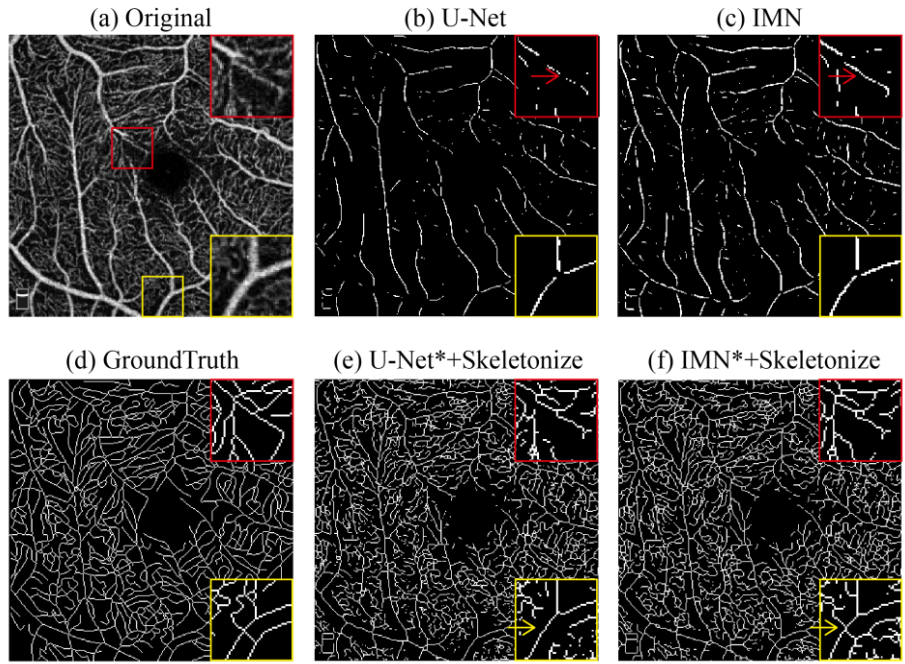

**Fig. 5.** An example of vessel skeleton segmentation on ROSE. The symbol * indicates the pixel-level segmentation model trained on OCTA-SS.

We tried another vessel skeleton segmentation strategy. We first used the model trained on OCTA-SS to segment the pixel-level vessels on the OCTA images of ROSE, and then used the method [26] to extract the vessel skeleton. The experimental results are shown in Figures 5(e)(f). This method achieves performance beyond the ground truth for manual labeling because more capillary skeletons are identified. Similarly, the vessel skeleton segmentation based on IMN has more vessel details than that based on U-Net (Figure 5, yellow arrows). Therefore, instead of directly using the neural network to segment the vessel skeleton, we recommend segmenting the pixel-level blood vessels first and then skeletonizing.

## 5    Conclusion

In this paper, we developed a novel end-to-end framework named IMN for retinal vessel segmentation in OCTA images. It contains an up-sampled encoding path and a down-sampled decoding path. This design is to get better detail retention capabilities, thereby more accurately segmenting thin-and-small vessels. The experiments on three open datasets indicate that the proposed method achieves the state-of-the-art performance of vessel segmentation in OCTA images. We also found that the IMN model trained on one FOV image can be used for another FOV image, and its pixel-level vessel segmentation results can be used for the extraction of vessel skeletons.